# The superconductivity mechanism in Nd-1111 iron-based superconductor doped by calcium

F. Shahbaz Tehrani, V. Daadmehr[*]

*Magnet & Superconducting Research Lab., Faculty of Physics & Chemistry, Alzahra University, Tehran 19938, Iran*

[*]Corresponding author:
Tel: (+98 21) 85692640 / (+98) 912608 9714
Fax: (+98 21) 88047861
E-mail: daadmehr@alzahra.ac.ir
URL: http:// staff.alzahra.ac.ir/daadmehr/
First author:
E-mail: tehrani66@gmail.com

**Acknowledgements**

The authors are grateful to Vice Chancellor Research and Technology of Alzahra University for financial supports.

**Abstract**

We described the effect of nonmagnetic impurity on the superconductivity behavior of the NdFeAsO$_{0.8}$F$_{0.2}$ iron-based superconductor. The resistivity measurement showed that the superconductivity suppressed upon increasing the low amounts of calcium impurity (x≤0.05). Also, the Tc decreased with the increase in the residual resistivity. Such behavior was qualitatively described by the Abrikosov-Gorkov theory and confirmed that these impurities act as scattering centers. For our samples, the exchange constant between the calcium and the conduction electron spins was estimated J$_{exc}$=|8| meV. Moreover, we presented the phase diagram of our synthesized samples for the various calcium dopings and found that according to increase of the calcium impurities and temperature decreasing of the spin-density wave (T$_{SDW}$), Fe ions arranged stripe antiferromagnetically at lower temperatures and also the superconducting transition temperature (T$_C$) declined. Based on our results and in agreement with the available theories as explained in the text, since the S++ state has not an effect on the impurity-doped samples, and the S± state that is attributed to the spin-fluctuations mechanism causes the superconducting suppression for low amounts of calcium. So, it confirms the role of the spin-fluctuations as a dominant pairing mechanism in our synthesized samples.

# 1. Introduction

For the first time, the discovery of superconductivity in LaFeAsO$_{1-x}$F$_x$ compounds (with T$_C$=26 K) by Y. Kamihara et al. [1] attracted the attention of physicists to the iron-based superconductors (FeSCs) as a new group of superconductors and introduced them as a candidate for another group of high-temperature superconductors. The FeSCs have similar characteristics to the cuprates such as the layered structure, the small coherence length, the dependence of the superconductivity transition temperature (T$_C$) upon the doping, and the unconventional pairing mechanism [2-6]. Also, they display the extraordinary physics due to the co-existence of the magnetism and superconductivity, and the multiband electronic structure [7,8]. But the superconducting mechanism of these compounds is one of the controversial issues and there is still no explicit answer to this question. Most FeSCs have a common phase diagram that there is often a structural transition from tetragonal to orthorhombic and a magnetic spin-density wave (SDW) state with stripe antiferromagnetic (AF) spin configuration [9-12]. From this one can conclude that they play and important role in the superconducting mechanism of FeSCs. However, the cause of the structural transition is still under debate and has been proposed that either spin [13-15] or ferro-orbital nematic ordering [16-18] is responsible for the occurrence of it. In the spin-nematic scenario, the structural transition is driven by magnetic fluctuations and the lattice symmetry breaks from C$_4$ to C$_2$ through the magnetic-elastic coupling to lift the degeneracy of the stripe-type antiferromagnetism [19-22]. In the orbital-nematic scenario, the ferro-orbital ordering causes the breaking of C$_4$ symmetry and the strong inter-orbital interactions lead to an unequal occupation of the d$_{xy}$ and d$_{yz}$ Fe orbitals [23-25].

Moreover, up to now, two superconducting mechanisms have been reported for the FeSCs: spin- and orbital-fluctuations. Based on the spin-fluctuations theory, fully-gapped sign-reversing s-wave (S±) state had been predicted [26, 27]. The origin of the spin-fluctuations is the intra-orbital nesting and the Coulomb interaction. Furthermore, the orbital-fluctuations generally originate from the inter-orbital nesting and the electro-phonon interactions due to the Fe-ion optical phonons [28-30]. The theories introduced that the electron-phonon coupling improves the orbital fluctuations at wave vectors (0, 0) and ($\pi$, 0)/(0, $\pi$), which correspond to ferro-orbital and antiferro-orbital fluctuations, respectively [31, 32]. Also, the antiferro-orbital fluctuations around ($\pi$, 0)/(0, $\pi$) competes with the antiferromagnetic spin-fluctuations around the same wave vectors [5]. When the spin-fluctuations overcome, the pairing state is S± as mentioned above, but when the orbital fluctuations dominate, the sign of the gap remains the same between the electron and hole pockets and is called the S++ state [33-35].

The theoretical models [33, 36-38] described that the S± state is very fragile to the nonmagnetic impurities, and only 1% nonmagnetic impurity with moderate scattering potential could completely suppress superconductivity and the S± pairing state, while for the S++ symmetry the $T_C$ does not show the effective variation to the nonmagnetic impurities [33]. So, an appropriate probe to determine the type of symmetry and pairing in the FeSCs is the substitution of nonmagnetic impurity. Consequently, to discover the superconducting mechanism, several experimental works have been carried out for the investigation of the impurity effects on the suppression of superconductivity in the FeSCs [39-44]. For example, the Zn ions which act as strong potential scattering centers in the LaFeAsO$_{1-x}$F$_x$ compound and suppress the superconductivity [45]. However, the conclusions remain discussable and the more experimental and theoretical works are needed to illuminate the impurity effects in the FeSCs that can help us to understand the superconducting mechanism. In our previous work, we studied the effects of $Ca^{2+}/Nd^{3+}$ substitution on polycrystalline $Nd_{1-x}Ca_xFeAsO_{0.8}F_{0.2}$ samples with $0 \leq x \leq 0.1$ [46]. We compared our experimental data and the results of the mentioned theories i.e. the spin- and orbital-fluctuations in pairing mechanism. The consistency of our experimental results and the theoretical reports based on the spin- and the orbital- fluctuations theories shows that these models play an important role in the pairing mechanism of the iron-based superconductors. Given this background, in this work, we focus on the investigation of calcium doping effects as a nonmagnetic impurity in the superconductivity mechanism of Nd-1111 iron-based superconductor. We study the calcium doping effects on the structural transition and SDW state and the relation between them to clarify the superconducting mechanism. We show that calcium doping can affect the spin- and orbital-fluctuations and change the competition between them. We hope that our experimental results and the available corresponding theories will help to understand the pairing symmetry and superconductivity mechanism in the FeSCs.

## 2. Experimental

We used one-step solid state reaction method for synthesizing the polycrystalline samples with the nominal compositions of $Nd_{1-x}Ca_xFeAsO_{0.8}F_{0.2}$ (x=0.0, 0.01, 0.025, 0.05, and 0.1), as indicated in Refs. [46,47]. The $NdFeAsO_{0.8}F_{0.2}$, $Nd_{0.99}Ca_{0.01}FeAsO_{0.8}F_{0.2}$, $Nd_{0.975}Ca_{0.025}FeAsO_{0.8}F_{0.2}$, $Nd_{0.95}Ca_{0.05}FeAsO_{0.8}F_{0.2}$ and $Nd_{0.9}Ca_{0.1}FeAsO_{0.8}F_{0.2}$ samples are labeled as Nd-1111, Nd-Ca0.01, Nd-Ca0.025, Nd-Ca0.05, and Nd-Ca0.1, respectively. The X-ray diffraction patterns (XRD) of our samples were accomplished using a PANalytical® PW3050/60 X-ray diffractometer with Cu Kα radiation (λ= 1.54056 Å) operated at 40 kV and 40 mA with a step size of 0.026°. We applied a four-probe technique for the superconductivity measurements using the 20K Closed Cycle Cryostat (QCS101), ZSP Cryogenics Technology. Also, we used a Lake Shore-325 temperature controller for measuring the temperature and the applied DC current (Lake Shore-120) was 10 mA and the voltage was measured with microvolt accuracy.

## 3. Results and discussion

Fig. 1 shows the XRD patterns of the $Nd_{1-x}Ca_xFeAsO_{0.8}F_{0.2}$ samples that the rather pure phase and the negligible impurities in each pattern indicate the good quality of our synthesized samples. All samples have the tetragonal structures with P4/nmm:2 space group that proved with Rietveld's analysis by employing the MAUD software. Further details of this analysis were explained in our previous work in Ref. [46]. The variation of lattice parameters "a, c" and the cell volume for the calcium contents is displayed in Fig. 2. All the data are normalized by the values of the undoped sample i.e. Nd-1111. As shown, the lattice parameters and the cell volume decrease upon increasing the calcium doping.

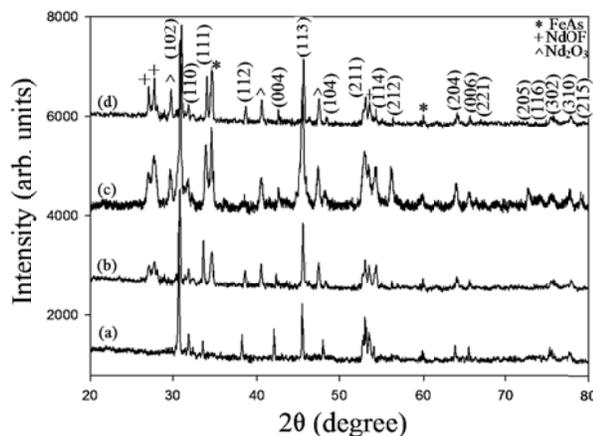

Fig. 1. The XRD patterns of our samples: (a) Nd-1111, (b) Nd-Ca0.01, (c) Nd-Ca0.025 and (d) Nd-Ca0.05

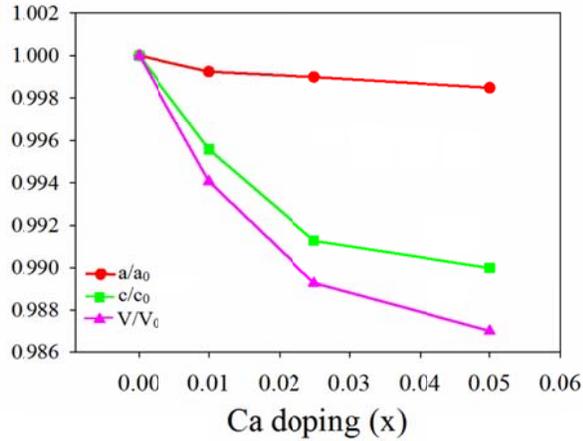

Fig.2. The variation of normalized lattice parameters and cell volume for the various calcium dopings

The decline of the lattice parameter "c" is more in comparison to the "a". Due to the small difference in the ionic radius of the calcium and neodymium ions, the reason for the reduction of the mentioned parameters is the change in other structural parameters such as bond lengths, bond angles, and the thickness of the layers. In our previous work, we explicitly explained the reason for the reduction of the lattice parameters, based on the variation of bond lengths, bond angles and the thickness of layers with the increase in the calcium content [46]. The temperature dependence of resistivity for our synthesized samples is shown in Figs. 3-6. As shown in these figures, the $T_C$ of our samples decreases by increasing the calcium doping and also, the superconductivity suppresses in the Nd-Ca0.05 sample. In our previous work [46], the dependence of $T_C$ and the As-Fe-As bond angles, the Fe-As bond length, the pnictogen height, and the lattice parameters for the various calcium contents have been discussed. We have understood that the $T_C$ decreased with increasing the distortion of FeAs4-tetrahedron from the regular value and also with decreasing the Fe-As bond length, the pnictogen height, and the lattice parameters with the increase in the calcium content. Our experimental data were consistent with the same results that had suggested theoretically based on the spin- and orbital-fluctuations theories (as described in Ref. [46]). Now, we want to know which type of spin- or orbital-fluctuations is dominated in our samples, so, we study the investigation of calcium impurity effects via the available theories.

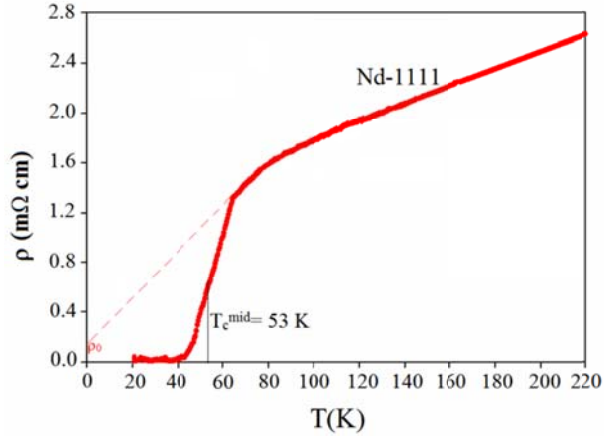

Fig. 3. Temperature dependence of resistivity for the synthesized Nd-1111 sample

**Suppression of superconductivity by nonmagnetic impurity-** From the X-ray diffraction data, we can see that calcium content could be easily placed into the neodymium sites of Nd-1111 sample up to 0.05 [46]. According to Abrikosov-Gorkov theory (AG), if the impurities act as strong pair breakers, the suppression of $T_C$ is associated with the impurity scattering rate define as $k_B \Delta T_C \approx \pi \hbar \tau_s \propto \rho_0$. We define the $\rho_0$ from the extrapolation of resistivity in the low temperature region before starting of the superconductivity state that is displayed for our synthesized samples by a dashed line in Figs. 3-6. Figs. 7(a) and (b) display that the values of $\rho_0$ and the normalized $T_C$ are found to increase and decrease, respectively, by increasing the calcium impurity. Also, the decrement of the normalized $T_C$ versus the calcium doping is linear with a good approximation. Furthermore, the dependence of $T_C$ and $\rho_0$ is shown in Fig.8, which describes the suppression of superconductivity in our synthesized samples with the increment of $\rho_0$. So, the substitution of calcium ions strongly destroys superconductivity and increases the residual resistivity ($\rho_0$) for our samples, which means that these impurities act as scattering centers.

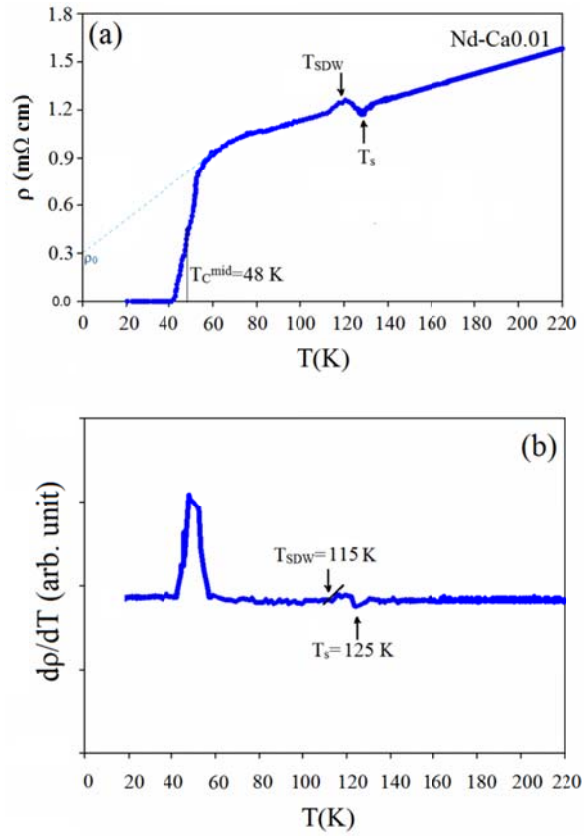

Fig. 4. Temperature dependence of (a) resistivity and (b) its derivate for the synthesized Nd-Ca0.01 sample

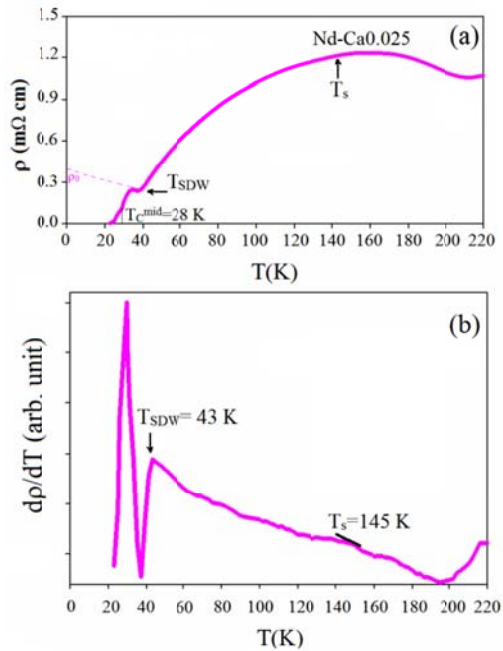

Fig. 5. Temperature dependence of (a) resistivity and (b) its derivate for the synthesized Nd-Ca0.025 sample

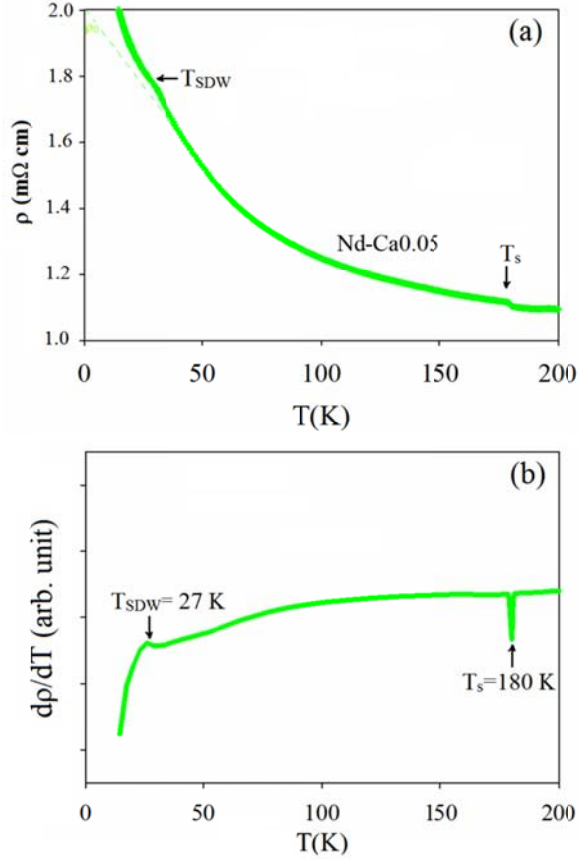

Fig. 6. Temperature dependence of (a) resistivity and (b) its derivate for the synthesized Nd-Ca0.05 sample

Based on the AG theory, in the presence of nonmagnetic impurities and isotropic exchange interaction, the variation of $T_C$ is given by the following expression [48-51]:

$$\ln[\frac{T_C}{T_{C0}}] = \Psi\left[\frac{1}{2} + (2\pi k_B T_C \tau_s)^{-1}\right] - \Psi\left(\frac{1}{2}\right) \quad (1)$$

with

$$\frac{1}{\tau_s} = 2\pi n_I N(0)(g-1)^2 J_{exc}^2 \, J(J+1) \quad (2)$$

Where $n_I$ the concentration of impurity ions, $T_{C0}$ is the transition temperature in the absence of doping, $\Psi$ is the digamma function, $N(0)$ is the density of states at the Fermi level. Also, g and J is the Lande g factor and the total angular momentum of the impurity ion, respectively. Likewise, $J_{exc}$ is the exchange constant between the impurity ion spin and the conduction electron spin. For the small concentration of nonmagnetic impurities, the $T_C$ decreases linearly with $n_I$ and the initial rate of depression of $T_C$ ($dT_C/dn_I$) is determined by the following equation [48, 51]:

$$\frac{dT_C}{dn_I}\Big|_{n_I \to 0} = \frac{dT_C}{dx}\Big|_{x \to 0} = -\frac{\pi^2 N(0) J_{exc}^2 (g-1)^2 J(J+1)}{2k_B} \quad (3)$$

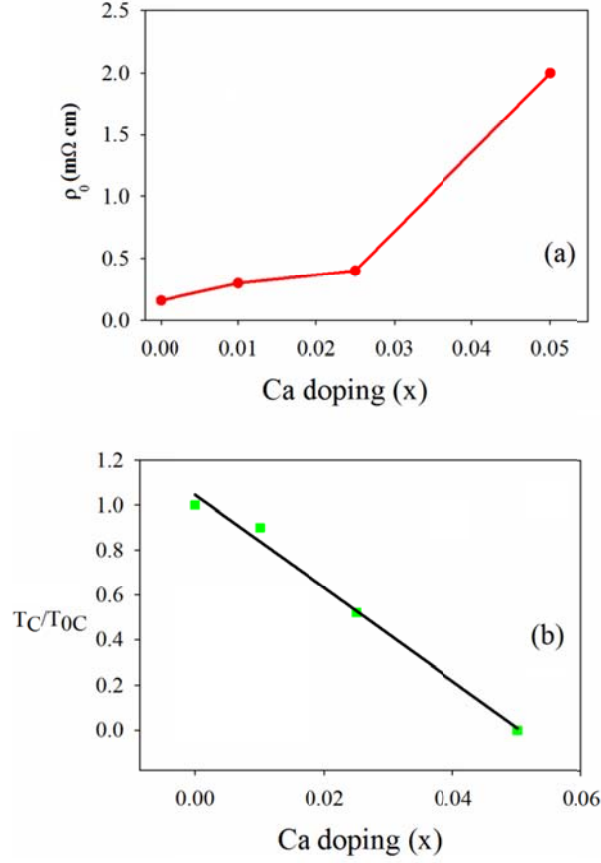

Fig. 7. The variation of (a) $\rho_0$ and (b) $T_C/T_{C0}$ as a function of calcium contents (x).

For $Nd_{1-x}Ca_xFeAsO_{0.8}F_{0.2}$ samples with one dopant atom per unit cell, we define $x=n_I$ where x is the fraction of calcium ions in our synthesized samples. The variation of $T_C$ with the concentration of calcium doping was shown in Fig. 7(a).

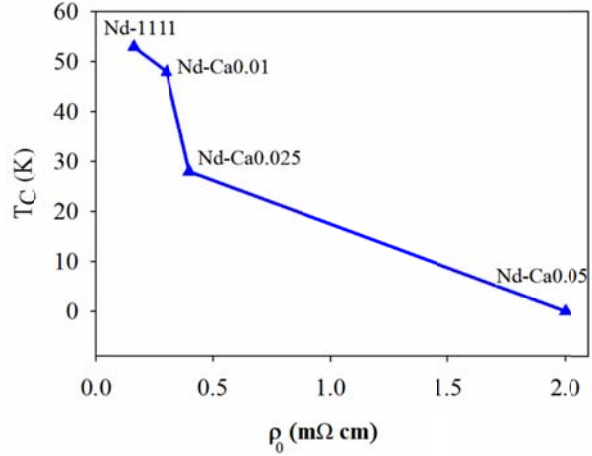

Fig.8. The dependence $T_C$ of versus $\rho_0$ for our synthesized samples

With the calculation of line slope (by using the experimental value), we have $\frac{dT_c}{dx}=$ -81.15 K/atom and so obtain from Eq. (3):

$$N(0)J_{exc}^2 \approx 70.9 \times 10^{-5} \left(\frac{eV.states}{atom\ spin}\right) \qquad (4)$$

We get a value of N(0) =10 states/eV atom spin from Ref. [52]. Therefore, the exchange constant ($J_{exc}$) between the calcium ion spin and the conduction electron spin is estimated |8| meV. This amount has the same order of magnitude that obtained for the 122-type of FeSCs and the conventional superconductors in Refs. [53,54] that obtained based on the ESR experiments [44]. The dependence result of $T_C$ ($\rho_0$) and based on the AG theory shows that calcium impurity ions behave similarly to the magnetic ions with $J_{exc}$=|8| meV in our samples. Consequently, the AG theory describes the competition effects of the magnetic and nonmagnetic impurities in the superconductivity destruction.

**Phase diagram of the synthesized samples-**As shown in Figs.4-6, there are other shoulders in the resistivity measurements of the calcium doped samples. These shoulders were usually attributed to the structural transition ($T_S$) and the SDW state ($T_{SDW}$) [55]. In 1111-type of FeSCs, the $T_{SDW}$ is smaller than $T_S$ [56], while for 122-type is the same and they move away from each other by substitution the doping and applying the pressure [57]. For more accurate, we calculate the derivate of resistivity and obtained the values of $T_S$ and $T_{SDW}$ for our synthesized samples. Fig. 9(a) displays the phase diagram of our samples and shows the suppression of the superconducting transition and the SDW state with increasing the calcium impurities, while the $T_S$ increases. With the increase of calcium impurities and the decrement of $T_{SDW}$, it can be described that the Fe ions arrange stripe AF at the lower temperatures and so $T_C$ declines. Moreover, Fig.9 (a) presents a similar trend of decreasing for the $T_{SDW}$ and $T_C$. As an important result, the spin-fluctuations have a direct and effective impact on the superconductivity mechanism of our samples. Likewise, as we described before based on the AG theory, the calcium ions act as scattering centers and hence, at the lower temperatures, the Fe ions arrange stripe AF. This means that nonmagnetic impurities can affect and suppress the $T_{SDW}$ and correspondingly the $T_C$. Based on the five-orbital model, S. onari et al. [33] theoretically had considered the effect of the local and low amount of impurity in FeSCs and found that the interband impurity scattering was boosted by the d-orbital degree of freedom. This signifies that the fully gapped sign-reversing S± state, which is attributed to the spin-fluctuations theory, is very fragile against impurities, while the S++ state due to the orbital-fluctuation is constant for all amount of impurity and not affect in the superconductivity suppression (for more detail see Fig. 9(b)). So the matching of our experimental phase diagram and the results of the above theoretical model confirms the role of the spin-fluctuations as a dominant pairing mechanism in our synthesized samples. In other words, the low amount of calcium impurity (x≤0.05) in our samples suppresses completely the superconductivity. Consequently and according to the mentioned theory, the S± symmetry and the spin-fluctuations are predominant against to the orbital-fluctuations in our samples. In the other work, J. Li et al. [58] described that the $T_C$ would be weakly suppressed by impurities in the S++ state, because of the following reasons: (1) Suppression of the orbital-fluctuations because of the violation of the orbital degeneracy near the impurities, (2) The strong localization effect, which the mean free path is comparable to the lattice spacing. Also, the ARPES evidence for S± symmetry, in contrast to the S++ state confirms the opinion that the superconducting and the SDW states can be suppressed by the same impurities [59-61]. This behavior has occurred in our samples. Since impurity has not an effect on the S++ state in the doped samples, and the S± state that is attributed to the spin-fluctuations mechanism causes the superconducting suppression for low amounts of the impurity, it can be concluded that the superconducting mechanism of our samples is the spin-fluctuations coupling. This result is in agreement with our previous work [46].

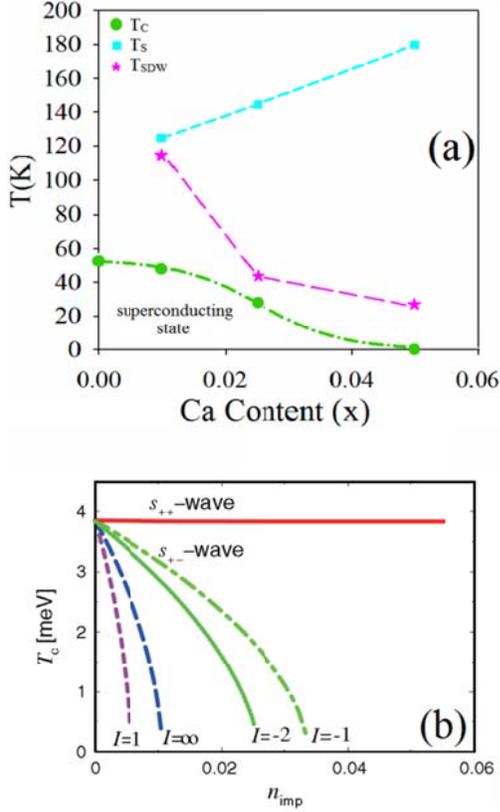

Fig. 9. (a) Phase Diagram of our synthesized samples , (b) Theoretical phase diagram for arbitrary nonmagnetic impurity that taken from Ref. [33]

## 4. Conclusions

To summarize, we have studied the superconducting behavior of the Nd-1111 iron-based superconductor that was doped by the calcium impurity. The samples were synthesized through the one-step solid state reaction method. We found that:

1. Superconductivity had been suppressed for the low amount of calcium impurity.
2. The $T_C$ decreased linearly with the increase of impurity and also $T_C$ reduced by increasing the residual resistivity. So, the calcium impurity acted as scattering centers and the Abrikosov-Gorkov theory was true for our samples. Accordingly, we estimated the exchange constant between calcium and conduction electron spins (Jexc=|8| meV).
3. In phase diagram of our synthesized samples, upon increasing the calcium impurity, the temperature of the spin-density wave ($T_{SDW}$) state declined, so at the lower temperatures, the Fe ions arranged stripe antiferromagnetically.
4. According to the previous results and the matching of our experimental- and the theoretical-phase diagram that can concluded from the available theoretical model, the role of the spin-fluctuations (the S± state) has been confirmed as a dominant pairing mechanism in our samples.